\documentclass[sigconf]{acmart}
\AtBeginDocument{%
  }

\setcopyright{acmlicensed}
\copyrightyear{2025}
\acmYear{2025}
\acmDOI{XXXXXXX.XXXXXXX}

\acmConference[MM '25]{Proceedings of the 33th ACM International Conference on Multimedia}{27 October - 31 October 2025}{Dublin, Ireland}
\acmISBN{978-1-4503-XXXX-X/2018/06}
\usepackage{multirow}
\usepackage{diagbox}

\acmSubmissionID{4143}

\begin{document}

\title{DMF2Mel: A Dynamic Multiscale Fusion Network for EEG-Driven Mel Spectrogram Reconstruction}

\thanks{
$^\star$Corrresponding author}


\author{Cunhang Fan}
\affiliation{%
  \institution{School of Computer Science and Technology}
  \institution{Anhui University}
  \city{Hefei}
  \country{China}}
\email{cunhang.fan@ahu.edu.cn}

\author{Sheng Zhang}
\affiliation{%
  \institution{School of Computer Science and Technology}
  \institution{Anhui University}
  \city{Hefei}
  \country{China}
}
\email{e23301156@ahu.edu.cn}

\author{Jingjing Zhang}
\affiliation{%
  \institution{School of Computer Science and Technology}
  \institution{Anhui University}
  \city{Hefei}
  \country{China}
}
\email{e22201067@stu.ahu.edu.cn}

\author{Enrui Liu}
\affiliation{%
  \institution{School of Computer Science and Technology}
  \institution{Anhui University}
  \city{Hefei}
  \country{China}
}
\email{e23201090@ahu.edu.cn}

\author{Xinhui Li}
\affiliation{%
  \institution{School of Computer Science and Technology}
  \institution{Anhui University}
  \city{Hefei}
  \country{China}
}
\email{xinhuili@ahu.edu.cn}

\author{Gangming Zhao}
\affiliation{%
  \institution{School of Computer Science and Technology}
  \institution{Anhui University}
  \city{Hefei}
  \country{China}
}
\email{gmzhao@connect.hku.hk}

\author{Zhao Lv{$^\star$}}
\affiliation{%
  \institution{School of Computer Science and Technology}
  \institution{Anhui University}
  \city{Hefei}
  \country{China}
}
\email{kjlz@ahu.edu.cn}

\renewcommand{\shortauthors}{Cunhang Fan et al.}

\begin{abstract}
Decoding speech from brain signals is a challenging research problem. Although existing technologies have made progress in reconstructing the mel spectrograms of auditory stimuli at the word
or letter level, there remain core challenges in the precise reconstruction of minute-level continuous imagined speech: traditional
models struggle to balance the efficiency of temporal dependency
modeling and information retention in long-sequence decoding.
To address this issue, this paper proposes the Dynamic Multiscale Fusion Network (DMF2Mel), which consists of four core components: the Dynamic Contrastive Feature Aggregation Module (DC-FAM), the Hierarchical Attention-Guided Multi-Scale Network (HAMS-Net), the SplineMap attention mechanism, and the bidirectional state space module (convMamba). Specifically, the DC-FAM separates speech-related "foreground features" from noisy "background features" through local convolution and global attention mechanisms, effectively suppressing interference and enhancing the representation of transient signals. HAMS-Net, based on the U-Net framework, achieves cross-scale fusion of high-level semantics and low-level details. The SplineMap attention mechanism integrates the Adaptive Gated Kolmogorov-Arnold Network (AGKAN) to combine global context modeling with spline-based local fitting. The convMamba captures long-range temporal dependencies with linear complexity and enhances nonlinear dynamic modeling capabilities. Results on
the SparrKULee dataset show that DMF2Mel achieves a Pearson
correlation coefficient of 0.074 in mel spectrogram reconstruction
for known subjects (a 48\% improvement over the baseline) and
0.048 for unknown subjects (a 35\% improvement over the baseline).Code is available at: https://github.com/fchest/DMF2Mel.
\end{abstract}



 \maketitle

\section{Introduction}
Decoding speech from brain activity has long been a goal in medicine and neuroscience. Brain-computer interfaces (BCIs) offer hope for patients who have lost the ability to speak due to brain injuries, strokes, or neurodegenerative diseases\cite{ref1,ref2,ref3,tankus2012structured}. Previous studies have successfully decoded speech features\cite{wandelt2024representation}, acoustic features\cite{akbari2019towards,moses2016neural,li2021human}, articulatory movements\cite{ref4}, and semantic information\cite{ref5,willett2023high} from intracranial recordings. However, while invasive devices have made significant progress in decoding basic linguistic features, they are not suitable for most patients\cite{wang2022open}. Therefore, decoding speech from non-invasive electroencephalogram (EEG) signals has become an important research direction. Non-invasive BCIs do not require surgical implantation of electrodes, making them safer, more acceptable, and applicable to a broader population\cite{krishna2020speech}. Nevertheless, challenges remain, including the weak and easily interfered nature of EEG signals\cite{graimann2010brain,fan2024dgsd}, as well as the difficulty in precisely extracting information due to the involvement of multiple brain regions in language processing\cite{fan2024msfnet}.

With the growth of non-invasive BCI technology, researchers' attempts to decode linguistic features from EEG signals have gradually deepened. Early work focused on low-level semantic classification, such as using multi-channel convolutional neural networks (MC-CNN) to identify the grammatical categories of words with implicit articulation (verbs/nouns)\cite{datta2021recognition}, achieving an accuracy of 85.7\%. A spatiotemporal model\cite{mahapatra2023eeg} combining one-dimensional convolutional neural networks (1D-CNN) and long short-term memory networks (LSTM) achieved an accuracy of over 85\% in character and digit classification tasks. These studies demonstrated the feasibility of EEG decoding, but their limitations lie in the closed vocabulary set and simple semantic hierarchy, making it difficult to handle complex language sequences in real-world scenarios. Subsequent research shifted to speech feature reconstruction, such as the NeuroTalk framework\cite{lee2023towards}, which used generative adversarial networks and speech synthesis technology to reconstruct words from articulatory and imagined EEG signals. However, these methods rely on small-scale datasets and are limited to word-level speech, with a significant gap from the long-term continuous speech decoding required for practical applications. Long-term imagined speech modeling needs to address noise interference in EEG signals, challenges in multi-scale feature integration, and difficulties in modeling strong temporal dependencies. Noise can easily mask critical transient features, and the cross-scale collaboration between local articulatory movements and global semantic representations in brain language processing is difficult to capture\cite{fan2025m3anet}\cite{fan2025listennet}. Traditional models also generally suffer from temporal information decay or computational efficiency issues in long-sequence decoding.

To overcome the above limitations, this paper proposes the Dynamic Multiscale Fusion Network (DMF2Mel) for EEG-Driven mel spectrogram reconstruction, which introduces a hierarchical architecture optimized through dual-branch feature extraction, multi-scale collaborative fusion, and deep temporal modeling. First, the dual-branch feature enhancement module consists of Dynamic Contrast Feature Aggregation (DC-FAM) and Hierarchical Attention-Guided Multi-Scale Network (HAMS-Net), which address local feature enhancement and cross-scale feature integration, respectively. DC-FAM separates speech-related "foreground features" from noisy "background features" by contrasting local convolution and global attention, effectively suppressing interference and enhancing transient signal representation. HAMS-Net, based on a U-Net architecture\cite{ronneberger2015u}, integrates top-down attention mechanisms to link high-level semantic information with low-level detail features, addressing the disconnection between local and global features in traditional models. Second, the multi-scale collaborative fusion module introduces a SplineMap attention mechanism, which dynamically balances "local precise fitting" and "global robust representation" through an Adaptive Gated Kolmogorov-Arnold Network (AGKAN), achieving efficient cross-branch feature fusion for EEG signals with non-stationarity and individual differences. Finally, the deep temporal modeling module employs a Convolution-Enhanced Bidirectional State Space Model (convMamba), which captures long-range temporal dependencies with linear complexity, while residual connections and feedforward networks enhance nonlinear dynamic modeling capabilities, addressing temporal misalignment and information decay in long-sequence decoding.

The main contributions of this paper can be summarized as follows:
\begin{itemize}
    \item  We propose a dual-branch feature enhancement architecture that includes a Dynamic Contrast Feature Aggregation Module and a Hierarchical Attention Network. This architecture separates speech features from noise, integrates multi-scale information, addresses the issue of local and global semantic disconnection, and enhances the effective representation of EEG signals.
    \item We designed a dynamic multi-scale fusion mechanism based on SplineMap attention. It balances local detail and global context modeling, adapts to EEG non-stationarity and individual differences, and boosts the robustness of cross-subject feature fusion.
    \item Experiments on the SparrKULee dataset demonstrate that DMF2Mel achieves state-of-the-art (SOTA) performance in mel spectrogram reconstruction, with a 48\% improvement over the baseline for held-out stories (Pearson correlation coefficient of 0.074) and a 35\% improvement for held-out subjects (0.048). Ablation studies validated the necessity of the proposed modules, providing an effective solution for non-invasive BCIs applications.
\end{itemize}

\section{Related Works}
The core challenges of long-distance decoding imagined speech from non-invasive EEG signals lie in effectively handling the low SNR, multi-scale feature coupling, and temporal dynamics of EEG signals. Early research focused on speech envelope reconstruction, capturing low-frequency amplitude variations to reflect speech rhythm. For example, the VLAAI network\cite{accou2023decoding} introduced nonlinear components and output context modules to enhance temporal dependency modeling. The FFT model\cite{piao2023happyquokka} used pre-layer normalization to optimize the Transformer structure for single-subject scenarios. WaveNet\cite{van2023decoding} improved long-sequence processing efficiency through non-causal dilated convolutions. These methods improved envelope decoding accuracy by enhancing network architectures but were limited by the singularity of low - frequency information, making it hard to retain high-frequency acoustic details of speech and showing insufficient robustness to individual differences and noise interference.

As research advances toward high-precision speech feature reconstruction, the mel spectrogram has become a core target due to its rich time-frequency information. SSM2Mel\cite{fan2025ssm2mel} leverages the linear complexity advantage of state space models (SSM) to process long-term EEG sequences, combining the S4-UNet structure to enhance local feature extraction. ConvConcatNet\cite{xu2024convconcatnet} integrates multi-channel spatial information through deep convolution and feature concatenation, supported by an envelope auxiliary training strategy to optimize spectral details. CAT-WaveNet\cite{li2024cross} employs a coarse-to-fine granularity strategy, using cross-attention mechanisms to progressively reconstruct envelopes, multi-band mel spectrograms, and amplitude spectra, thereby strengthening cross-modal associations. These approaches attempt to improve spectral reconstruction quality through multi-stage generation or cross-modal fusion. However, they still face the issue of multi-scale feature fragmentation—specifically, the collaboration mechanism between local transient features and global semantic representations in EEG has not been effectively modeled\cite{yan2024darnet}. Traditional convolutional or attention modules also have limited noise suppression capabilities for weak signals, leading to poor generalization in cross-subject scenarios, especially for unseen held-out subjects, where models struggle to adapt to individual neural activity differences.

Current approaches focus separately on modeling the temporal structure of speech envelopes and reconstructing the details of mel spectrograms, yet no unified framework has emerged that balances noise robustness, multi-scale feature integration, and temporal dynamics. Designing adaptive feature processing modules to achieve efficient mapping from local neural activity to global speech representations is the key to advancing EEG-driven speech decoding technology toward practical applications.

\section{Overall Architecture}

\begin{figure*}
    \centering
    \includegraphics[width=0.9\textwidth]{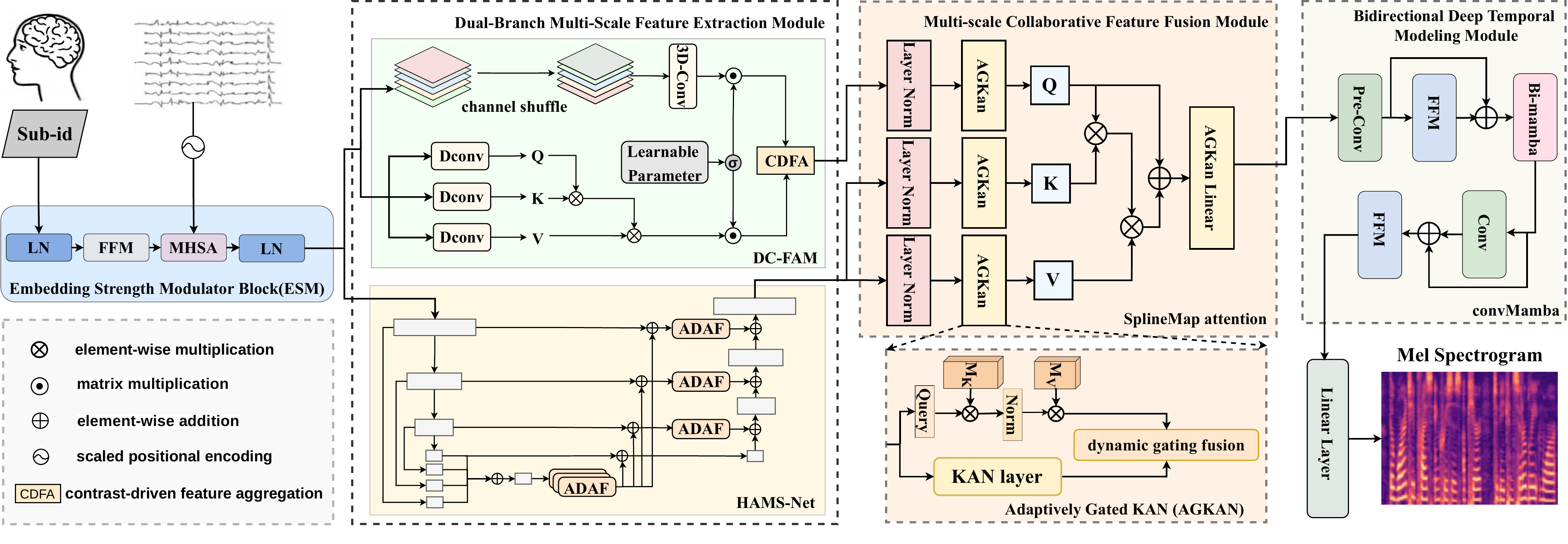}
    \caption{The overall framework of the DMF2Mel model, which consists of Embedding Strength Modulator (ESM), Dynamic Contrastive Feature Aggregation Module (DC-FAM), Hierarchical Attention-Guided Multi-Scale Network (HAMS-Net), SplineMap Attention Mechanism, and Bidirectional State Space Module (convMamba). }
    \label{fig:enter-label}
\end{figure*}

In this section, we introduce the Dynamic Multiscale Fusion Network (DMF2Mel), as illustrated in Figure 1.

\subsection{Embedding Strength Modulator}
The Embedding Strength Modulator (ESM) serves as the foundational module for personalized feature adaptation, dynamically integrating subject-specific information to address inter-individual variability in EEG signals. Based on prior work\cite{fan2025ssm2mel}, ESM encodes subject IDs into embedding vectors that are fused with raw EEG features through multi-head attention, feedforward networks, and layer normalization, enabling the model to capture personalized neural-physiological differences. This adaptive modulation adjusts feature representations to align with each subject’s unique neural responses, providing a tailored input foundation for the subsequent dual-branch feature extraction module. By leveraging subject ID embeddings to modulate EEG features, ESM enhances the model’s sensitivity to individual neural patterns, a critical step for improving generalization in non-invasive speech decoding.

\subsection{Dual-Branch Multi-Scale Feature Extraction Module}
To address the issues of local feature masking caused by noise interference and cross-scale feature disconnection in EEG signals, DMF2Mel has designed a dual-branch feature enhancement module. 

\subsubsection{Dynamic Contrastive Feature Aggregation Module (DC-FAM)}
In EEG signal processing, the effective neural activity features directly related to speech imagination ("foreground" features) are often masked by noise and physiological artifacts ("background" noise), making it difficult to extract critical transient signals. The Dynamic Contrast Feature Aggregation Module (DC-FAM), designed to address this issue, is built on the CAFM module\cite{hu2024hybrid} and adopts a dual-branch structure. Inspired by the ConDSeg framework in medical image segmentation\cite{lei2024condseg}, DC-FAM provides a novel technical approach tailored to the characteristics of neural signals for EEG denoising and feature enhancement.By leveraging the Contrast-Driven Feature Aggregation (CDFA) mechanism, it explicitly distinguishes "foreground features" (speech-related effective signals) from "background noise" (interfering signals). This mechanism not only suppresses noise interference but also enhances the representation of effective neural activities.

The DC-FAM module consists of a local convolutional branch and a global attention branch. The local convolutional branch focuses on extracting transient details of "foreground" features, while the global attention branch aims to suppress long-range interferences from "background" noise.The local convolutional branch adjusts the channel dimension through convolution and uses depthwise separable convolution and channel shuffling to enhance spatial correlation, highlighting transient signals related to speech. It also introduces learnable weight parameters to improve the model's adaptability to individual-specific features.The global attention branch generates query (Q), key (K), and value (V) matrices using 1×1 convolution and 3×3 depthwise separable convolution, computes attention scores to capture long-range dependencies in the temporal dimension, and enhances noise suppression through learnable weight amplification of the output.

 Then, the outputs of the two branches are fused through the CDFA module.Given the local features \( F_{conv} \)  from the convolutional branch and the global features \( F_{atten} \) from the attention branch, it first generates a contrastive attention map:
\begin{equation}
A_{fg} = softmax(W_{fg}F_{conv})
\end{equation}
\begin{equation}
A_{bg} = softmax(W_{bg}F_{atten})
\end{equation}
Here, \( W_{fg} \) and \( W_{bg} \) are learnable projection matrices. Local attention \( A_{fg} \) focuses on transient features of EEG signals, while global attention \( A_{bg} \) models semantic correlations across time segments and suppresses background noise through long-range dependencies.
For the output of the convolutional branch, local neighborhood feature blocks are extracted through an Unfold operation.
\begin{equation}
V_{unfold} = Unfold(F_{conv} )
\end{equation}
Then,local and global features are contrastively fused through matrix multiplication.
\begin{equation}
O_{fg}=\sum_{i=1}^{k^{2} }  (A_{fg}[:,:,:,i,:]\odot V_{unfold}  )
\end{equation}

\begin{equation}
O_{bg}=\sum_{i=1}^{k^{2} }  (A_{bg}[:,:,:,i,:]\odot V_{unfold}  )
\end{equation}
Here,\( O_{fg} \) and \( O_{bg} \) respectively achieve local feature weighting and global context injection.Finally, the outputs of the dual branches \( O_{fg} \) and \( O_{bg} \) are reconstructed to the original spatial dimensions.
\begin{equation}
O = Reshape(O_{fg} + O_{bg}  )
\end{equation}

The DC-FAM module suppresses background noise and enhances foreground transient features via dual branches and the CDFA mechanism, thereby improving the capture of key EEG features in complex noisy environments. The clean local features it generates complement the cross-scale semantic features of HAMS-Net, providing structured inputs for multiscale fusion. This effectively addresses the issues in traditional models where noise obscures details and transient signals are difficult to extract, thereby enhancing the quality of EEG feature representation.

\subsubsection{Hierarchical Attention-Guided Multi-Scale Network (HAMS-Net)}
Inspired by the top-down regulatory mechanism in human speech synthesis, where high-level semantic information guides low-level speech generation\cite{gazzaley2005top} , we designed the Hierarchical Attention-Guided Multi-Scale Network (HAMS-Net). Based on the U-Net architecture, this network employs an explicit semantic guidance mechanism to optimize local feature generation using global semantic information, addressing the issue of disconnection between local details and high-level semantics in traditional models\cite{fan2025cross}. Its multi-scale feature extraction strategy, combined with hierarchical attention modules, enables deep cross-level feature fusion, allowing global semantic information to directly participate in and optimize the generation of local features.

Specifically, the encoding process extracts multi-scale features from EEG signals through multiple convolutional layers, capturing higher-level semantic information as the resolution decreases. These features are fused into a comprehensive representation, which is first enhanced by multiple layers of the Adaptive Dual-Attention Feedback (ADAF) module.

The ADAF module dynamically adjusts the spatial (PAM) and channel (CAM) attention\cite{woo2018cbam} branches through adaptive weighting and feedback (as shown in Figure 2). It first extracts features through convolution and then applies the PAM and CAM modules to generate spatial dependency features and channel screening features. Different from the fixed-weight design of traditional dual-attention modules, ADAF introduces two independent learnable parameters w1 and w2, which are activated by the Sigmoid function to obtain dynamic weights \( \alpha \)  and \( \beta \), respectively weighting and adjusting the outputs of PAM and CAM. The feedback mechanism uses the features from the previous layer to optimize the feature extraction of the current layer, which is a key improvement over the traditional Dual-Attention module, enabling the network to dynamically adapt during training. The formula is:

\begin{equation}
F_{ADAF}  = \alpha \cdot PAM(x) + \beta \cdot CAM(x) + Feedback(x_{prev} )
\end{equation}
where \( x_{prev } \) is the output of the previous ADAF layer.

In the decoding stage, the multi-scale features optimized by the ADAF module are upsampled through interpolation and concatenated layer-by-layer with the features of different scales in the encoding stage to form a composite representation interweaving semantics and details. At this time, the ADAF module intervenes again to enhance the features. Finally, through multiple layers of progressive decoding and attention optimization, an output matching the input resolution is generated, resulting in the final decoding result.

\begin{figure}[htbp]
    \centering
    \includegraphics[width=\linewidth]{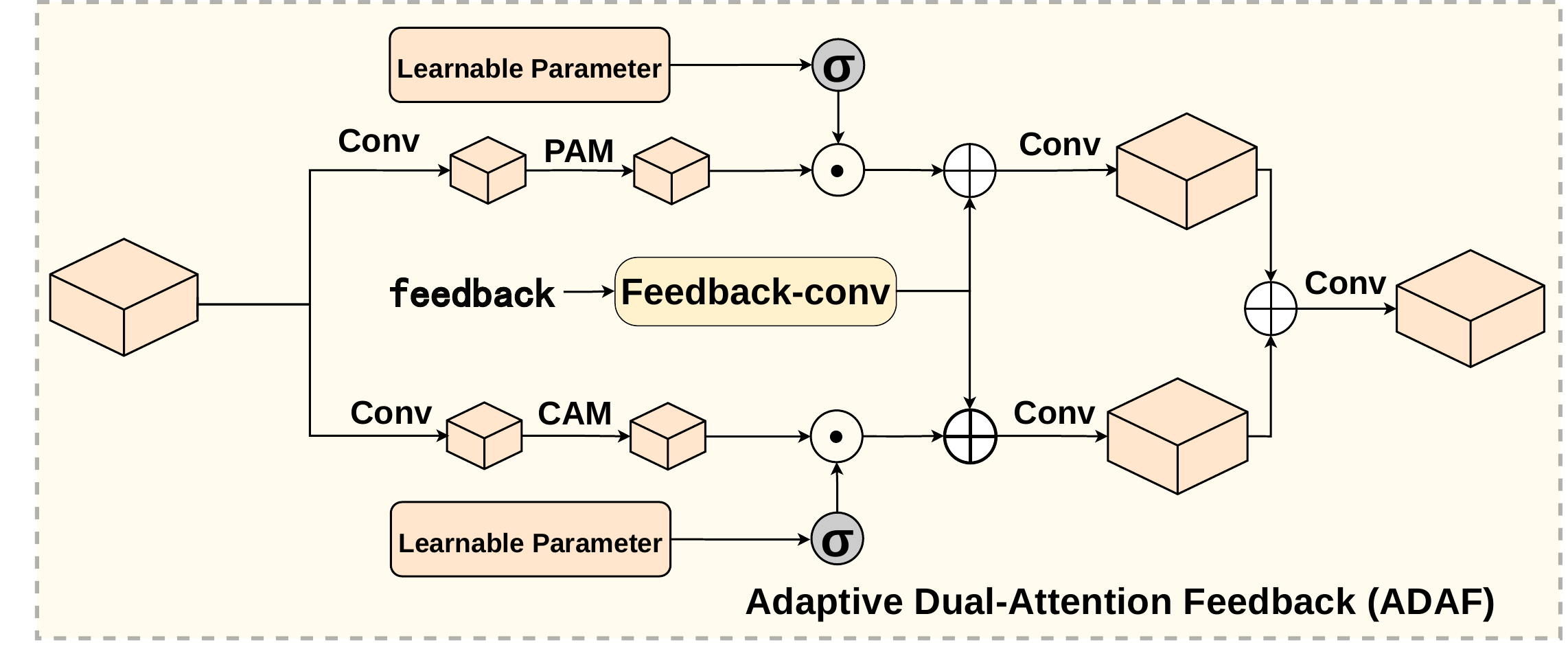} 
    \caption{The components of the ADAF block in HAMS-Net. } 
    \label{fig:result} 
\end{figure}

The hierarchical attention-guided feature fusion mechanism addresses the disconnection between global and local semantics in traditional models by reconstructing multi-scale EEG features layer-by-layer from coarse-grained semantics to fine-grained details. Explicit cross-scale interactions enhance feature mutual information, effectively capturing the coordinated activity patterns of multiple brain regions during language processing. This creates a semantically consistent feature space for subsequent fusion, achieving organic unity between low-level details and high-level semantics, and laying a critical foundation for multi-scale modeling of complex EEG signals.

\subsection{Multi-scale Collaborative Feature Fusion Module}
To address the non-stationarity and inter-individual variability of EEG signals, we propose a feature fusion mechanism based on multi-layer SplineMap Attention. This module effectively integrates differentiated features extracted by multi-branch networks. Innovatively, it combines the global context modeling of traditional attention mechanisms with the local fitting advantages of spline functions, constructing a dynamic feature balancing framework.

Standard multi-head attention has limitations in EEG-driven imagined speech decoding. Its linear projection layer struggles to capture the highly nonlinear mapping between neural activity and speech representations\cite{yang2025reynolds}, while its fixed basis function combinations fail to address the non-stationary noise and individual differences in EEG signals. Although traditional convolutional or fully connected layers enhance detail modeling through local feature extraction, they face contradictions in modeling long-term temporal dependencies: the former cannot capture cross-time-step semantic associations, while the latter suffers from feature distortion due to noise accumulation.

The Kolmogorov-Arnold Network (KAN)\cite{liu2024kan} demonstrates potential for time-series modeling through non-linear fitting with spline functions\cite{lee2024hippo,xu2024kan4drift}. However, its single-variable optimization strategy has inherent limitations: it is sensitive to local noise and lacks global context modeling, making it difficult to capture cross-band long-range semantic dependencies in speech generation\cite{zhou2024kan}. To address these issues, this paper proposes the Adaptive Gated Kolmogorov-Arnold Network (AGKAN) (as shown in Figure 1). AGKAN adopts a dual-path heterogeneous architecture to combine strengths: it retains KAN's local feature parsing capability while introducing an external attention mechanism with global compressive sensing properties. This mechanism captures cross-time-step sequence patterns through low-rank projection. Additionally, a dynamic gating network is designed to adaptively adjust the weights (gating values  \( g \in [0,1] \)) of the dual paths via differentiable learning, achieving a balance between "local precise fitting" and "global robust representation."

Based on this, AGKAN further constructs the SplineMap attention module, which adaptively transforms the core components of traditional attention mechanisms: in the attention mapping layer, learnable nonlinear spline basis functions are used to replace linear projections, enhancing transient EEG features through adaptive spline approximation; in the output layer, an individual-adaptive spectral reconstruction mechanism is built using AGKAN's piecewise continuous function space. This hybrid architecture retains the global context modeling capability of attention mechanisms. Meanwhile, leveraging AGKAN's distribution-adaptive characteristics, it effectively balances the non-stationarity of EEG signals and individual differences, significantly enhancing the model's ability to fuse "local neural details and global speech semantics" across scales.

 Specifically, \( H_1 \) is multi-scale features extracted by the hierarchical attention architecture (HAMS-Net), and \( D_1 \) is features extracted by the dual-branch feature aggregation module (DC-FAM). 

Initial Fusion: First, interact \( D_1 \) with \( H_1 \)'s features by setting \( D_1 \) as the query vector to obtain the initial fused feature \( F_1 \):
\begin{equation}
F_{1} =SplineMapAttention(D_{1} ,H_{1},H_{1})
\end{equation}

Progressive Interaction: Fuse intermediate results \( F_1 \) with features from \( H_1 \) and \( D_1 \) step-by-step. At each step, dynamically adjust the query, key, and value vectors based on the previous step's output to incrementally enhance and integrate features.

This progressive interaction ensures the final output feature \( F_t \)integrates features from \( H_1 \) and \( D_1 \), capturing both global context and local details. This design leverages the strong modeling capabilities of SplineMap Attention and boosts feature expressiveness through step-by-step interaction, providing richer feature representations for subsequent decoding tasks.

\subsection{Bidirectional Deep Temporal Modeling Module}
Efficient modeling of long-sequence temporal dependencies in EEG signals, following the dynamic fusion of multi-scale features, is crucial for enhancing the accuracy of mel spectrogram reconstruction.To address this, we propose the Convolutional Enhanced Bidirectional State Space Model (convMamba) module. The core of this module lies in the Bidirectional State Space Model (BiMamba)\cite{liu2024vmamba}, which captures global dependencies in sequences while addressing the issues of information decay and computational inefficiency commonly encountered in traditional models during long-sequence decoding\cite{fan2025seeing}.

In the convMamba block, features are first transformed through a convolutional layer for linear transformation, which enhances their expressiveness by mapping them to the target dimension. Subsequently, the BiMamba module processes the embedded features bidirectionally, capturing comprehensive context in the sequence with linear complexity. The core design is as follows:
First, the forward state update.
\begin{equation}
h_{t} = \bar{A}h_{t-1} + \bar{B}x_{t}  
\end{equation}
\begin{equation}
y_{t} = Ch_{t}    
\end{equation}
In this formula, \( \bar{A} \) is the learnable state transition equation, \( \bar{B} \) is the input matrix, C is the output mapping matrix, \( h_t \) is the hidden state, \( x_t \) is the input, and \( y_t \) is the output at time step t.
The backward process performs reverse state transitions using the same parameter matrices.
\begin{equation}
g_{t} = \bar{A}g_{t+1} + \bar{B}x_{t}  
\end{equation}
\begin{equation}
y_{t} = Cg_{t}    
\end{equation}
The bidirectional outputs are fused through temporally aligned concatenation, enabling the simultaneous capture of both forward and backward contextual information.To enhance the model's nonlinear expression, feedforward networks (FFN) are added both before and after the BiMamba module. This module effectively processes long sequences while fully preserving the complex temporal dynamics of EEG signals, thus enabling the precise reconstruction of time-domain continuous mel spectrograms.

Finally, the block's output is mapped through a linear layer to produce the final feature representation.

\section{Loss Function}

The loss function combines correlation measurement, sparsity constraints, and contrastive learning to optimize the model's ability to generate Mel spectrograms. The total loss function is composed of three parts: correlation measurement, sparsity constraints, and contrastive learning.The total loss function is:
\begin{equation}
\mathcal{L}_{total} =  \mathcal{L}_{p} +\lambda  \mathcal{L}_{1} + \beta \mathcal{L}_{InfoNCE}
\end{equation}
where \( \lambda  \) and \( \beta  \) are adjustable hyperparameters.

Pearson loss \(\mathcal{L}_p\) aligns temporal structures by measuring covariance-standard deviation ratios between decoded signals and target spectrograms, driving the model to learn speech-like temporal patterns.

\begin{equation}
\mathcal{L}_{p}=1-\frac{cov(\hat{Y},Y )}{\sigma _{\hat{Y} } \sigma _{Y }}  = 1-\frac{ {\textstyle \sum_{i=1}^{T}(\hat{Y_{i} }-\mu _{\hat{Y} }  )({Y_{i} }-\mu _{{Y} }  )} }{\sqrt{ {\textstyle \sum_{i=1}^{T}} (\hat{Y_{i} }-\mu _{\hat{Y} } )^{2} } \sqrt{{\textstyle \sum_{i=1}^{T}} ({Y_{i} }-\mu _{{Y} }  )^{2} } }
\end{equation}
Where \( \mu \) denotes the mean, \( \sigma \) represents the standard deviation, and \( cov \) represents the covariance,where \( \hat{Y} _{i} \)   represents the predicted value of the mel spectrogram, \( \ Y_{i} \) represents the actual value,

The  \(\mathcal{L}_1\) enforces sparsity through mean absolute error minimization, improving detail accuracy while suppressing outliers.
\begin{equation}
\mathcal{L}_{1} =\frac{1}{n}  {\textstyle \sum_{i=1}^{n}\left | \hat{Y}_{i}-Y_{i}    \right | } 
\end{equation}

The \(\mathcal{L}_{InfoNCE}\), symmetrically formulated as \(L_{\text{eeg} \to \text{audio}}\) and \(L_{\text{audio} \to \text{eeg}}\), maximizes cross-modal mutual information via cosine-similarity-based contrastive learning, strengthening latent EEG-speech associations.
\begin{equation}
\mathcal{L}_{InfoNCE} = \frac{1}{2}[\mathcal{L}_{eeg\longrightarrow audio}+\mathcal{L}_{audio\longrightarrow eeg}]
\end{equation}
\begin{equation}
\mathcal{L}_{eeg\longrightarrow audio} = -log\frac{exp(s(\hat{Y}_{i}, {Y}_{i})/\tau )}{ {\textstyle \sum_{j=1}^{N}} exp(s(\hat{Y}_{i}, {Y}_{j})/\tau )}
\end{equation}
\begin{equation}
\mathcal{L}_{audio\longrightarrow eeg} = -log\frac{exp(s({Y}_{i}, \hat{Y}_{i})/\tau )}{ {\textstyle \sum_{j=1}^{N}} exp(s({Y}_{i}, \hat{Y}_{j})/\tau )} 
\end{equation}
Where  \( s(\cdot   ) \) denotes the cosine similarity function, \( \tau \)  is the temperature coefficient, and N is the number of samples in a batch. This loss enhances the model's ability to capture cross-modal semantic associations by maximizing the mutual information between positive sample pairs.

\section{Experimental Setup}

\subsection{Datasets}
We utilize the dataset from the ICASSP 2024 Auditory EEG Challenge, SparrKULee\cite{bollens2023sparrkulee}, which includes EEG data from 85 Dutch-speaking participants with normal hearing. Each subject undergoes 8 to 10 trials lasting approximately 15 minutes, listening to stories told in fluent Flemish (Belgian Dutch). We apply multichannel Wiener filtering to the raw data to remove artifacts, then re-reference the EEG signals to a common average, and downsample the EEG signals to 64 Hz. We divided the dataset as follows: The training set includes 70 subjects, and the test set contains the same 70 subjects' responses to audio stimuli not present in the training set, which is referred to as "held-out stories." Additionally, the test set includes responses from 15 subjects not in the training set to the audio stimuli from the training set, which is referred to as "held-out subjects".

\begin{table*}[htbp]
\centering
\caption{Performance Comparison in Mel Spectrogram and Speech Envelope Reconstruction.Except for Sea-Wave* results from its original paper, the remaining methods’ (VLAAI, HappyQuokka, SSM2Mel) results are re-implemented in this study.} 
\begin{tabular}{c|ccc|ccc}
\hline
\multirow{2}{*}{Methods} & \multicolumn{3}{c|}{Mel spectrogram Reconvery}         & \multicolumn{3}{c}{Speech Envelope Recovery}           \\ \cline{2-7} 
                        & Held-out Stories & Held-out Subjects & Score           & Held-out Stories & Held-out Subjects & Score           \\ \hline
VLAAI\cite{accou2023decoding}                   & 0.05             & 0.0355                 & 0.0452               & 0.1527            & 0.1008                 & 0.1354               \\
Sea-Wave*\cite{van2023decoding}                & -                & -                 & -               & 0.1741           & 0.1123            & 0.1535          \\
HappyQuokka\cite{piao2023happyquokka}             & 0.052            & 0.038             & 0.0473          & 0.185            & 0.109             & 0.1597           \\
SSM2Mel\cite{fan2025ssm2mel}                & 0.066            & 0.041             & 0.0577          & 0.208            & 0.116             & 0.1773          \\ \hline
\textbf{DMF2Mel(ours)}          & \textbf{0.074}   & \textbf{0.048}    & \textbf{0.0653} & \textbf{0.223}   & \textbf{0.128}    & \textbf{0.1913} \\ \hline
\end{tabular}
\end{table*}

\subsection{Implementation Details}
We implement our proposed model using PyTorch. We train our model for 1000 epochs using the Adam optimizer with an initial learning rate of 0.0005. We employ the StepLR scheduler, which automatically decreases the learning rate by a factor of 0.9 every 50 epochs. During the training process, we utilize 5-second signal segments, randomly cropped from each EEG/speech mel spectrogram segment, to ensure stable training. For inference, we divide the input signal into several 5-second segments, process each segment individually, and then concatenate the outputs to form the complete mel spectrogram. 

\subsection{Evaluation Metrics}
In this paper, we primarily use the Pearson correlation (Pearson r) as the evaluation metric for our reconstruction task. When calculating the reconstruction score, we use a weighted sum of the Pearson correlations averaged over a set of held-out stories (\( S_{1}  \) , containing 70 known subjects) and a set of held-out subjects (\( S_{2}  \), containing 15 unknown subjects), computed as:
\begin{equation}
score = \frac{2}{3} \sum_{s\in S_{1} }^{} \frac{Pearson\, r_{s} }{\left | S_{1}  \right | } +\frac{1}{3} \sum_{s\in S_{2} }^{} \frac{Pearson\, r_{s} }{\left | S_{2}  \right | }
\end{equation}

\section{Results}

\subsection{Comparison with Baselines}
Table I summarizes the results of the baseline model and our proposed model in reconstructing speech mel spectrograms. We first compared our model with three state-of-the-art models (VLAAI, HappyQuokka, and SSM2Mel) on the "held-out stories" dataset. The results show that our model performs excellently in the case of seen subjects, achieving a Pearson correlation coefficient of 0.074, which represents a 48\% improvement over the baseline model. Specifically, compared to VLAAI, HappyQuokka,and SSM2Mel, our method shows increases in the Pearson correlation coefficient of 0.024, 0.016,and 0.008, respectively, further proving the effectiveness of our approach. This significant performance improvement is attributed to the innovative architecture and optimization strategy of our model, which fully reflects its advantages in feature extraction and modeling.

\begin{figure}[htbp]
    \centering
    \includegraphics[width=\linewidth]{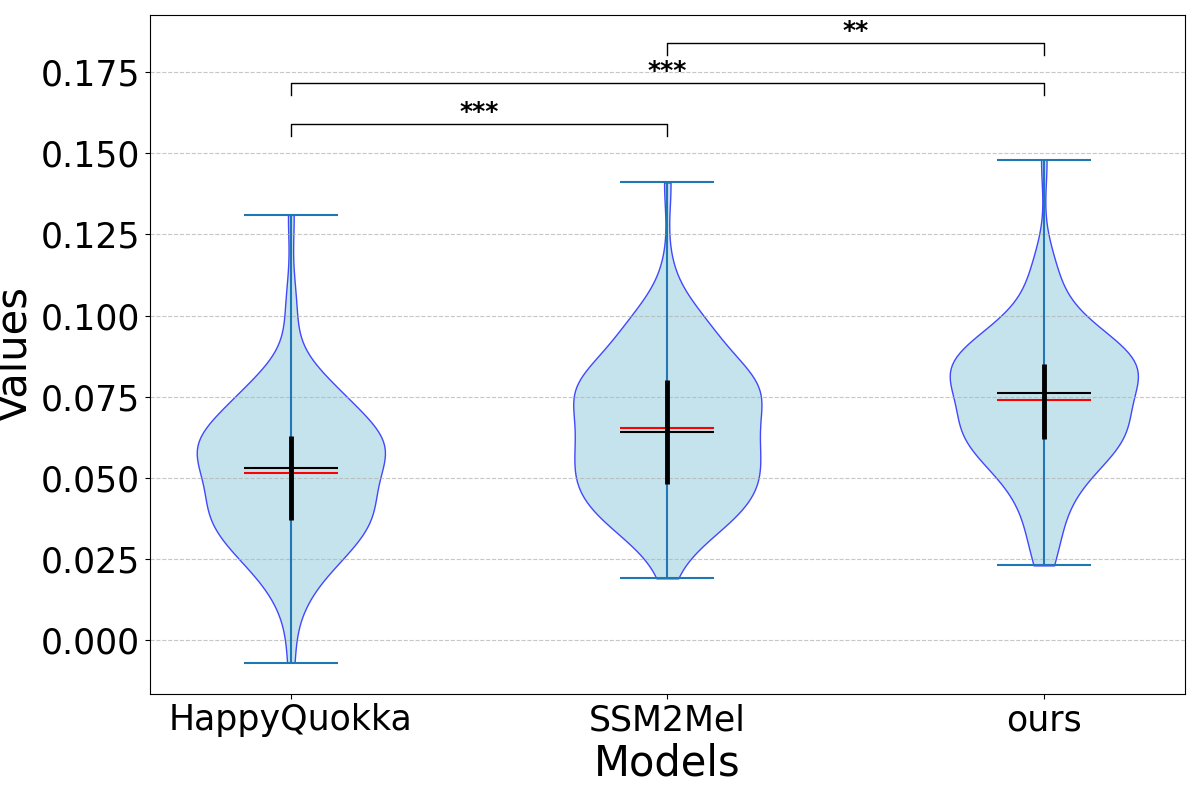} 
    \caption{Violin plot of the mel spectrogram test results for three models on the "held-out stories" dataset across subjects 1–85.  (**: p-value < 0.01, ***: p-value < 0.001)} 
    \label{fig:result} 
\end{figure}

To verify the generalization ability of our model, we conducted reconstruction tests on 15 unseen subjects from the "held-out subjects" dataset. We calculated the score of our model by combining the performance from both data splits and compared it with the models VLAAI, HappyQuokka and SSM2Mel. The results show that our model achieves a reconstruction similarity of 0.048 for the 15 unseen subjects, higher than the respective scores of 0.0125, 0.01 and 0.007 from VLAAI, HappyQuokka and SSM2Mel. The final score, calculated through a weighted average of performance across two datasets, demonstrates a significant advantage of our model in the mel spectrogram reconstruction task. Our model achieved a composite score of 0.0653, significantly outperforming VLAAI (0.0452), HappyQuokka (0.0473), and SSM2Mel (0.0577). This formula-based evaluation system systematically highlights the superior generalization capabilities and performance of our model in processing complex brain signals, showcasing its leading advantages across both specific generalization scenarios and overall performance.

\begin{figure}[htbp]
    \centering
    \includegraphics[width=\linewidth]{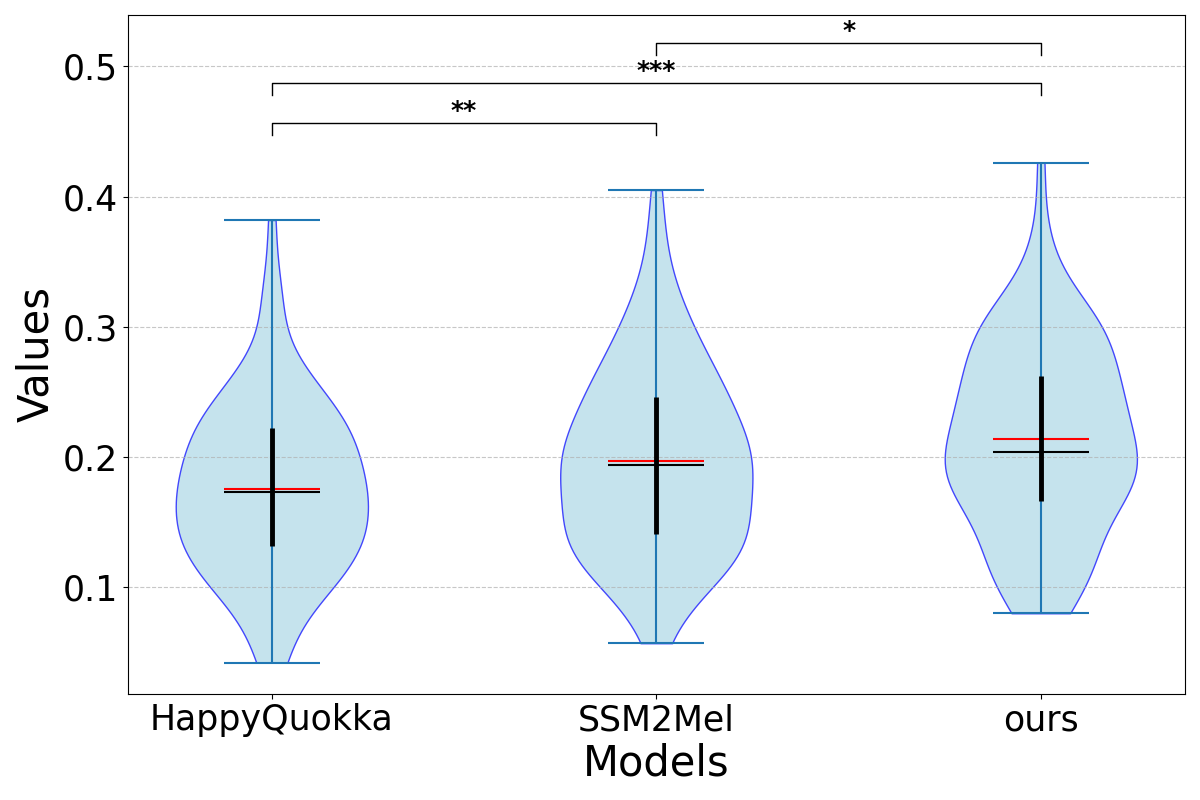} 
    \caption{Violin plot of the speech envelope test results for three models on the "held-out stories" dataset across subjects 1–85.  (*: p-value < 0.05,**: p-value < 0.01, ***: p-value < 0.001)} 
    \label{fig:result} 
\end{figure}

\begin{table}[]
\centering
\caption{Ablation Study on The Impact of Network Modules} 
\begin{tabular}{ccc}
\hline
Model         & Pearson correlation & \begin{tabular}[c]{@{}c@{}}Performance\\ improvement (\%)\end{tabular} \\ \hline
w/o ESM                 & 0.069              & -6.8                                                                 \\
w/o DC-FAM              & 0.040              & -45.9                                                                \\
w/o ADAF                & 0.059              & -20.3                                                                \\
w/o HAMS                & 0.056              & -24.3                                                                \\
w/o SplineMap Attention & 0.054              & -27.0                                                                \\
w/o convMamba           & 0.048              & -35.1                                                                \\ \hline
\textbf{DMF2Mel(ours)}           & \textbf{0.074}     & \textbf{-}                                                           \\ \hline
\end{tabular}
\end{table}

To further demonstrate the superiority of the proposed model, we compared the performance of VLAAI, Sea-Wave, HappyQuokka, and SSM2Mel on the task of speech envelope reconstruction using the same dataset, as shown in Table 1. The experimental results indicate that DMF2Mel consistently outperformed the baseline models across different evaluation scenarios. Specifically, in the "Held-out Stories" scenario, DMF2Mel achieved a Pearson correlation coefficient of 0.223 for speech envelope reconstruction. In the "Held-out Subjects" scenario, despite untrained individual neural activity variations, the model achieved a reconstruction coefficient of 0.128. From a comprehensive evaluation perspective, based on our score weighting rule, DMF2Mel obtained a high score of 0.1913 by integrating its performance in both scenarios, significantly outperforming the comparison models in all metrics. These results not only validate the leading performance of DMF2Mel in speech envelope reconstruction but also highlight its robustness in handling complex noise environments and individual differences, providing strong support for the practical application of non-invasive brain-computer interfaces.

To further demonstrate the robustness of our model, we compared the mel spectrogram reconstruction performance of DMF2Mel with two baseline models on the “held-out stories” dataset of subjects 1-85, and visualized the distribution of Pearson correlation coefficients for each subject using a violin plot (Figure 2). The results show that the median Pearson correlation coefficient for mel spectrogram reconstruction by the DMF2Mel model reached 0.077, significantly higher than HappyQuokka (0.053, p < 0.001) and SSM2Mel (0.064, p < 0.01). Additionally, Figure 3 presents the speech envelope reconstruction results of our model, which also exhibit superior performance. With the narrowest interquartile range, the model shows a concentrated performance distribution and minimal individual differences. This result confirms that DMF2Mel, through its dynamic feature focusing and multi-scale fusion mechanism, can effectively suppress noise and stably adapt to EEG signal characteristics of different subjects, showing strong robustness to individual differences and complex noisy environments.

\subsection{Ablation Study}

The ablation study results (Table 2) demonstrate that the full model (with a Pearson correlation coefficient of 0.074) significantly outperforms all module-removed versions in mel spectrogram reconstruction, verifying the necessity of multi-component collaborative design: Removing the Embedding Strength Modulator (ESM) caused a 6.8\% performance drop (0.069), confirming its value in cross-individual feature alignment. The removal of the Dynamic Contrastive Feature Aggregation Module (DC-FAM) led to the largest drop of 45.9\% (0.040), as it suppresses noise through foreground-background decoupling. Notably, the fusion module formed by DC-FAM and the Hierarchical Attention-Guided Multi-Scale Network (HAMS) via SplineMap attention becomes ineffective when either module is removed. HAMS removal caused a 24.3\% drop (0.056), disconnecting high-level semantics from low-level details. Removing the Adaptive Dual-Attention Feedback Module (ADAF) resulted in a 20.3\% drop (0.059), weakening multi-scale feature optimization. Replacing SplineMap attention with simple concatenation caused a 27.0\% drop (0.054), losing dynamic local-global balance. convMamba removal led to a 35.1\% drop (0.048), disrupting long-sequence temporal dependency modeling. The collaborative design of all modules is central to achieving optimal noise suppression, cross-scale fusion, and temporal modeling.

\begin{table}[]
\centering
\caption{Ablation Study on the Numbers of Modules} 
\begin{tabular}{c|ccccc}
\hline
\multirow{2}{*}{Model} & \multicolumn{5}{c}{Model Numbers} \\ \cline{2-6} 
                            & 1     & 2              & 4              & 6              & 8     \\ \hline
convMamba                   & 0.070 & \textbf{0.074} & 0.057          & -              & -     \\
DC-FAM                      & 0.067 & 0.066          & \textbf{0.074} & 0.060          & 0.055 \\
HAMS-Net                    & 0.068 & 0.066          & 0.071          & \textbf{0.074} & 0.061 \\
SplineMap Attention         & 0.063 & 0.067          & 0.066          & \textbf{0.074} & -     \\ \hline
\textbf{DMF2Mel(ours)}               & \multicolumn{5}{c}{\textbf{best:0.074}}                          \\ \hline
\end{tabular}
\end{table}

\begin{table}[]
\centering
\caption{Ablation Study on the Hyperparameter Selection in Loss Functions}
\begin{tabular}{c|lcccc}
\hline
\multirow{2}{*}{loss function} & \multicolumn{5}{c}{Hyperparameter Value} \\ \cline{2-6} 
                               & 0      & 0.1    & 0.3    & 0.5   & 1     \\ \hline
$\mathcal{L}_{1}$                              & 0.062  & 0.067  & 0.065  & \textbf{0.074} & 0.069 \\
$\mathcal{L}_{\text{info}} $                          & 0.07   & \textbf{0.074}  & 0.069  & 0.066 & 0.064 \\ \hline
\end{tabular}
\end{table}

Table 3 focuses on the impact of core module quantity configurations on model performance, taking the optimal combination of "2 convMambas, 4 DC-FAMs, 6 HAMS-Nets, and 6 SplineMap Attentions" as the benchmark, and keeping other modules at their optimal quantities when ablating one module. The results show that the convMamba module achieves the best Pearson correlation coefficient when the quantity is 2, decreases to 0.057 when increased to 4, and has no valid results due to overfitting with 6 or more. The DC-FAM module performs best with 4 modules, and the coefficients decrease to 0.066, 0.060, and 0.055 when reduced to 2 or increased to 6 and 8, respectively. The HAMS-Net module reaches a peak when there are 6 modules, and both too few and too many will reduce the integration ability of global semantics and local details; the SplineMap Attention module achieves the best dynamic multi-scale fusion with 6 modules. Overall, the results indicate that the performance of each module depends not only on their presence but also on achieving collaborative optimization through quantity configuration. The optimal quantity combination enables the model to form complementary functions in noise suppression, cross-scale fusion, and temporal modeling. Any deviation from the optimal quantity of any module will break this balance, leading to a decline in feature representation ability, and verifying the key impact of module quantity collaboration in the dynamic multiscale fusion architecture on decoding performance.

To verify the necessity of multi-component synergy in composite loss functions, our approach involved a stepwise parameter tuning strategy within the composite loss framework, where we fixed the optimal value of one component and adjusted the other, we conducted ablation analysis on \(\mathcal{L}_{1}\) sparsity loss and bidirectional contrastive loss (\(\mathcal{L}_{info}\)). Table 4 shows that when  \(\mathcal{L}_{1}\) constraint is applied, the reconstruction similarity peaks at 0.074 when the weight \( \lambda \) is set to 0.5, while over-constraint (\( \lambda \)=1) leads to loss of details. \(\mathcal{L}_{info}\) achieves the same peak value of 0.074 when \( \beta \) is set to 0.1, validating its role in feature distinguishability. Notably, when combining \(\mathcal{L}_{1}\) (\( \lambda \)=0.5) and \(\mathcal{L}_{info}\) (\( \beta \)=0.1), the composite loss function maintains a peak similarity of 0.074. The experimental results confirm that the combination of Pearson correlation loss, \(\mathcal{L}_{1}\) sparsity constraint, and bidirectional information noise contrastive estimation is key to improving model performance.

\section{Conclusion and Future Work}

This paper introduces the Dynamic Multiscale Fusion Network (DMF2Mel), addressing the challenge of balancing the efficiency of temporal dependency modeling and information retention in long sequences for traditional models. The proposed hierarchical architecture employs dual-branch feature extraction to enhance local transient details and global semantic context, alongside a multiscale fusion mechanism and bidirectional temporal modeling to handle EEG non-stationarity and individual differences, integrating dynamic feature focusing, multiscale fusion, and deep temporal modeling to enhance adaptability.  Experiments on the SparrKULee dataset demonstrate that DMF2Mel achieves state-of-the-art performance in mel spectrogram reconstruction, with significant improvements over baselines in both known and unknown subject scenarios, showcasing robust cross-subject generalization. Future work will target the core limitation of long-duration imagined speech decoding accuracy by developing adaptive temporal modeling architectures that better capture neural dynamics in continuous discourse.


\bibliographystyle{ACM-Reference-Format}
\bibliography{ref}

\end{document}